\documentclass[epj]{svjour}
\usepackage{amssymb}
\usepackage{graphicx}
\usepackage{dcolumn}
\usepackage{bm}
\usepackage{amsmath}
\usepackage{color}

\begin{document}
\title{Retrieving intracycle interference in angular-resolved laser-assisted
photoemission from argon}
%\subtitle{Do you have a subtitle?\\ If so, write it here}
\author{Johan Hummert\inst{1} \and Markus Kubin\inst{1,2} \and 
Sebasti\'{a}n D. L\'{o}pez\inst{3} \and Marc J. J. Vrakking\inst{1} \and Oleg
Kornilov\inst{1} \and Diego G. Arb\'{o}\inst{3}
% \thanks is optional - remove next line if not needed
%\thanks{\emph{Present address:} Insert the address here if needed}%
}                     % Do not remove
%
%\offprints{}          % Insert a name or remove this line
%
\institute{$^1$Max-Born-Institute, Max-Born-Stra{\ss}e 2A, 12489 Berlin,
Germany\\
$^2$Institute for Methods and Instrumentation for Synchrotron Radiation Research, Helmholtz-Zentrum Berlin für Materialien und Energie GmbH, Albert-Einstein-Strasse 15, 12489 Berlin, Germany\\
$^3$Institute for Astronomy and Space Physics IAFE (CONICET-UBA) CC 67, Suc. 28,
C1428ZAA, Buenos Aires, Argentina}
\date{Received: date / Revised version: date}
% The correct dates will be entered by Springer
%
\abstract{
We report on a combined experimental and theoretical study of XUV ionization
of atomic argon in the presence of a near-infrared laser field.
The resulting energy- and angle- resolved photoemission spectra
have been described in the literature as interferences among different 
photoionization trajectories. Electron trajectories stemming from different
optical laser cycles give rise to \textit{intercycle} interference energy and
result in new peaks known as sidebands. These sidebands are modulated by a
coarse grained (gross) structure coming from the \textit{intracycle}
interference of the electron trajectories born during the same optical cycle. 
We calculate the photoelectron emission by solving the time-dependent
Schr\"{o}dinger equation \textit{ab initio} and within the
continuum-distorted wave strong field approximation. In order
to compare with the experimental data we average the calculated energy-angle
probability distributions over the experimental focal volume. This averaging
procedure washes out the intracycle interference pattern, which we experimentally
and theoretically (numerically) recover by subtracting two averaged
distributions for slightly different near-infrared laser field intensities.
 % end of PACS codes
} %end of abstract
\maketitle
\section{Introduction}
\label{intro}

Generation of XUV pulses via high order harmonics opens new routes for
time-resolved spectroscopy with unprecedented time resolution. In combination
with near-infrared or visible (NIR/vis) laser light such XUV pulses allow for
control of electron motion on ultrashort timescales. The most readily observed
effect in such experiments is the laser-assisted photoelectric effect (LAPE), where
the NIR/vis light controls trajectories of free electrons created by the short
XUV pulses. For XUV pulses longer than the control laser cycle $T_{L}$,
the photoelectron energy spectrum shows a main line
associated with the absorption of one XUV photon accompanied by sideband (SB) lines.
The equally spaced SBs with separation $\hbar \omega _{L}$ are associated with
additional exchange of few NIR photons through absorption and stimulated emission
processes. The sidebands have been observed in laser-assisted ionization of gases,
liquids and solids and can be used to extract information on the XUV pulse duration,
control laser intensity, and time delay between the two pulses \cite%
{Itatani02,Fruehling09,Wickenhauser06,Duesterer13}.

For sufficiently high intensity of the control NIR/vis pulse another class
of features appears in photoemission spectra, which stem from the intracycle
interferences. For linearly polarized pulses they are manifested in modulations
of photoelectron angular distributions. This effect has been established for
ionization of Ne and other noble gases at free-electron laser facilities (FEL)
and compared with calculations \cite{Meyer10,Meyer12,Radcliffe12}.
However, to the best of our knowledge, a detailed analysis of
features in angular- and energy-resolved photoelectron spectra has not been
performed. We fill this gap and present a combined experimental and theoretical
study of intracycle and intercycle interferences in LAPE from Ar atoms.

An accurate theoretical description of the LAPE process must be based on quantum
mechanical concepts, i.e., on \textit{ab initio} solutions of the time dependent
Sch\"{o}dinger equation (TDSE) for the atomic system in the presence of the
electric field of two laser pulses. However, the precise calculation of the response of a rare gas
atom presents considerable difficulties. Often, the numerical solution of the
TDSE is employed, which relies on the single-active-electron (SAE) approximation,
with model potentials that permit one to reproduce the bound
state spectrum of the multi-electron atom with a satisfactory accuracy \cite%
{Nandor99,Muller99}. Models based on a time-dependent distorted wave theory,
like the strong field approximation (SFA) and the Coulomb-Volkov
approximation (CVA), have also been extensively employed to study LAPE (see for
example \cite{Kazansky10a,Kazansky10b,Bivona10}). To gain physical insight it is often
useful to compare full numerical results with qualitative predictions.
For example, the simple man's model \cite{Schins94}
has been very useful for the case of above-threshold ionization (ATI) by
one-color lasers where it is possible to describe the photoelectron spectra
as the interplay of intra- and intercycle interferences of 
electrons' trajectories \cite{Arbo10a,Arbo10b}.
On equal footing, the photoelectron spectra of
LAPE has recently been treated as an interference problem in the time domain \cite%
{Gramajo16,Gramajo17,Gramajo18}. Electron trajectories stemming from
different optical laser cycles give rise to \textit{intercycle} interference,
which results in new peaks, known as sidebands, modulated by a coarse-grained
structure coming from the \textit{intracycle} interference of the electron
trajectories born during the same optical cycle \cite%
{Gramajo16,Gramajo17,Gramajo18}.

In this paper, a detailed experimental and theoretical study of energy and
angle-resolved sideband spectra are presented.
XUV pulses from HHG are filtered by a time-delay-compensating monochromator
\cite{Eckstein15} and combined with the fundamental femtosecond NIR pulses.
Owing to precise synchronization of the laser pulses this experimental scheme
does not suffer from time jitter experienced in FEL-based experiments
\cite{Duesterer13}. The small bandwidth of the XUV pulses from the
monochromator can be used to study photoelectron sidebands which are
free from interference effects from adjacent harmonic orders.
This setup is hence an ideal tool for the investigation of sideband
generation and the related physics of strong-field and multiphoton nonlinear
effects which occur in LAPE. In combination with the velocity map imaging
spectrometer \cite{Eppink97}, angular distributions can be measured,
which render additional information on the photoionization process,
such as the partial wave character of the emitted photoelectrons or 
relative ionization phases.

We also perform theoretical calculations solving the full TDSE \textit{ab initio}
and employing the SFA to reproduce the experimental measurements.
We analyze angle-resolved photoelectron distributions of atomic argon in the case
where the duration of the XUV pulse $\tau _{X}$ is of the order of or larger
than the laser cycle period, i.e., $\tau _{X}\gtrsim T_{L}$. The role
of the NIR laser field in the XUV photoionization is threefold: (a) due to
Stark effect, it shifts the energy of the
continuum states of the atom down by the ponderomotive energy $U_{p}$, (b)
several NIR photons can be absorbed or emitted in the course of the
ionization process giving rise to SBs (or intercycle contributions), and (c)
it is responsible for intracycle modulations of the SBs in the photoelectron
angle-resolved spectrum. For the latter, the interfering electron trajectories
within the same optical cycle give rise to a well-determined modulation
pattern encoding information of the ionization process in the subfemtosecond
time scale. A similar behavior has been observed by J-W. Geng \textit{et al.}
in \cite{Geng13} for single ionization of He by considering two XUV
attosecond pulses separated by the duration of the laser cycle period.
In the experiments the intensity of the NIR field varies across the laser focus
leading to ``washing out'' of the interference features,
which are very sensitive to the field strength \cite{Kubin13}.
To compare theoretical calculations with the experimental results
we implement averaging of the calculated spectra over the focal volume.

The paper is organized as follows. In Sec. \ref{sec:theory} we describe 
the two different methods of calculating angle-resolved photoelectron
spectra for the case of laser assisted XUV ionization:
by solving the TDSE \textit{ab initio} and by making use of the SFA.
We also describe the calculation method to simulate the averaging
over the focal volume. In Sec. \ref{experiment} we describe
the experimental setup to measure the velocity-map images of atomic argon
ionized by XUV pulses in presence of the fundamental NIR laser
pulse and present the experimental results. In Sec. \ref{results}, we
present the theoretical results obtained solving the TDSE and within the SFA,
compare them with the measurements, and discuss the details 
of the sideband gross structure.
Concluding remarks are presented in Sec. \ref{conclusions}.

\section{\label{sec:theory}Methods of calculation}

We solve the problem of atomic ionization by an XUV pulse in the presence of
an NIR laser field with both fields linearly polarized along the $\hat{z}$ direction.
We use atomic units in this section, except when otherwise stated. The TDSE in the
single active electron approximation reads%
\begin{equation}
i\frac{\partial }{\partial t}\left\vert \psi (t)\right\rangle =H\left\vert
\psi (t)\right\rangle ,  \label{TDSE}
\end{equation}%
where the Hamiltonian of the system within the dipole approximation in the
length gauge is expressed as%
\begin{equation}
H=\frac{\vec{p}^{2}}{2}+V(r)+\vec{r}\cdot \vec{F}_{X}(t)+\vec{r}\cdot \vec{F}%
_{L}(t).  \label{hamiltonian}
\end{equation}%
The first term in Eq. (\ref{hamiltonian}) corresponds to the active electron
kinetic energy, the second term is the potential energy of the active
electron due to the Coulomb interaction with the core, and the last two
terms correspond to the interaction of the atom with the electric fields 
of the XUV pulse $\vec{F}_{X}(t)$ and the NIR pulse $\vec{F}_{L}(t)$.

As a consequence of the interaction between the atom and the XUV and NIR pulses,
the bound electron in the initial atomic state $|\phi _{i}\rangle $
is emitted into the final unperturbed state $|\phi _{f}\rangle $ with energy 
$E=k^{2}/2$ and angle $\theta $ with respect to the polarization axis $\hat{z%
}$. The angle- and energy-resolved continuum photoelectron distribution
can be calculated as 
\begin{equation}
\frac{dP}{\sin \theta dEd\theta }\mathbf{=}2\pi \sqrt{2E}\left\vert
T_{if}\right\vert ^{2},  \label{P}
\end{equation}%
where $T_{if}$ is the T-matrix element corresponding to the transition $\phi
_{i}\rightarrow \phi _{f}$. Eq. (\ref{P}) uses the cylindrical symmetry
of the problem around the polarization axis.

\subsection{Time-Dependent Schr\"{o}dinger Equation}

We employ the generalized pseudo-spectral method to numerically solve the
TDSE in the length gauge of the dipole approximation in the SAE
approximation \cite{Tong97,Tong00,Tong05}. 
Within this method the discretization of the radial coordinate is optimized
to incorporate the Coulomb singularity. Additionally suitable quadrature
methods allow for a stable long-time evolution using a split-operator
representation of the time-evolution operator.
Both bound as well as unbound parts of the wave function
$|\psi (t)\rangle $ can be accurately represented. The atomic potential $V(r)$
is modelled as the sum of the asymptotic Coulomb potential, $V(r)=-1/r$, and
a short-range potential accounting for the influence of the ionic core of Ar$%
^{+}$. Its parameters are chosen to reproduce the ionization potential $%
I_{p}=15.76$ eV and the energies of lower excited bound states \cite%
{Muller98}. Propagation of the wavefunction starts from the initial 3$p$
ground state orbitals $\varphi _{3p0}$ since the ionization from the $m=0$
orbital, aligned along the laser polarization axis, strongly dominates over $%
m=-1,1$ in the resulting spectrum. Due to the cylindrical symmetry
in a linearly polarized laser field, the magnetic quantum number is
conserved during the time evolution. After integrating over the duration
of the pulse, the wavefunction is projected onto eigenstates
$|k,\ell \rangle $ of the field-free atomic Hamiltonian with positive
eigenenergy $E=k^{2}/2$ and orbital quantum number $\ell $
to determine the transition amplitudes $T_{if}$
(see Refs.~\cite{Schoeller86,Messiah65,Dionis97}), i.e.,%
\begin{equation}
T_{if}=\frac{1}{\sqrt{4\pi k}}\sum_{\ell }e^{i\delta _{\ell }(p)}\ \sqrt{2l+1%
}P_{\ell }(\cos \theta )\left\langle p,\ell \right. \left\vert \psi
(t_{f})\right\rangle .  \label{coulomb}
\end{equation}%
In Eq.~(\ref{coulomb}), $\delta _{\ell }(p)$ is the momentum-dependent
atomic phase shift, $\theta $ is the angle between the electron momentum $%
\vec{k}$ and the polarization direction $\hat{z}$, and $P_{\ell }$ is the
Legendre polynomial of degree $\ell $. In order to minimize unphysical
reflections of the wave function at the boundary of the calculation box,
the length of the calculation box was chosen to be 1200 a.u.\ ($\sim 65$ nm)
and the maximum angular momentum considered was $\ell _{\max }=300$.

\subsection{Strong field approximation}

Within the time-dependent distorted wave theory, the transition amplitude in
the post form and length gauge is expressed as
\begin{equation}
T_{if}=-i\int_{-\infty }^{+\infty }dt\left\langle \chi _{f}^{-}(\vec{r}%
,t)\right\vert \left[ \vec{r}\cdot \vec{F}_{X}(t)+\vec{r}\cdot \vec{F}_{L}(t)%
\right] \left\vert \phi _{i}(\vec{r},t)\right\rangle \,  \label{Tif}
\end{equation}%
where $\phi _{i}(\vec{r},t)=\varphi _{i}(\vec{r})\,\mathrm{e}^{iI_{p}t}$ is
the initial atomic state with ionization potential $I_{p}$ and $\chi
_{f}^{-}(\vec{r},t)$ is the distorted final state \cite{Macri03}.
As the SFA neglects the Coulomb distortion in the final channel,
the distorted final wave function can be written as 
$\chi _{f}^{-}(\vec{r},t)=\chi ^{V}(\vec{r},t),\,$ where 
\begin{equation}
\chi ^{V}(\vec{r},t)=\,\frac{\exp {[i}\left( \vec{k}+\vec{A}{(t)}\right)
\cdot \vec{r}{]}}{(2\pi )^{3/2}}\exp {\left[ \frac{i}{2}\int_{t}^{\infty
}dt^{\prime }\left( \vec{k}+\vec{A}(t^{\prime })\right) ^{2}\right] }\,
\label{Volkov}
\end{equation}%
is the length-gauge Volkov state \cite{Volkov35} and 
$\vec{A}(t)=-\int_{-\infty }^{t}\vec{F}(t^{\prime })dt^{\prime }$
is the vector potential due to the combined electric field
\begin{equation}
\vec{F}(t)=\vec{F}_{L}(t)+\vec{F}_{X}(t).  \label{total-field}
\end{equation}%
For the sake of simplicity, we consider ionization of atomic argon which we
model as a hydrogen-like atom with effective charge $Z_{eff}=\sqrt{%
2n^{2}I_{P}}$, where $n$ is the principal quantum number of the initial state.
This effective charge ensures the ionization potential to be
taken into account properly and, consequently, the intra- and intercycle
fringes in electron spectra to be situated at the correct energy values
compared to full TDSE simulations. For electron energies far away from
the ionization threshold, the SFA is well known to give a good description of LAPE.
The good agreement between SFA
and TDSE ionization simulations in LAPE arises from the fact that the main
contribution to the ionization comes from on-shell transitions,
contrarily to NIR ionization, where off-shell transitions are dominant.
In other words, the absorption and emission of photons in LAPE take place
over real states: Firstly, an energetic XUV photon is absorbed followed
by the absorption and emission of NIR laser photons into the continuum.

\subsection{Average over the focal volume}

Laser intensity variation over the focal volume implies variations
of the LAPE spectra. Following the work of Posthumus for a single
color laser \cite{Posthumus09,Augst91}, we characterize the intensity
profile by a Lorentzian distribution along the laser propagation direction
$\hat{\zeta}$ and a Gaussian distribution along the radial direction
$\hat{\rho}$ in the perpendicular plane, with cylindrical symmetry around the
$\zeta$ axis. In the center of the waist (where the focusing effect is maximum),
i.e., at $\zeta =\rho =0$, the intensity of the XUV and laser fields are
$I_{X0}$ and $I_{L0},$ respectively. Within these assumptions,
the intensity distribution for the XUV and NIR laser beams are

\begin{equation}
I_{X,L}=\frac{I_{X0,L0}}{1+\left( \frac{\zeta}{\zeta _{X,L}}\right) ^{2}}\exp
\left\{ -\frac{2\rho ^{2}}{w_{X,L}^{2}\left[ 1+\left( \frac{\zeta }{\zeta
_{X,L}}\right) ^{2}\right] }\right\} ,  \label{IXL}
\end{equation}%
where $w_{X,L}$ is the waist radius, and $\zeta _{X,L}$ is the Rayleigh
range of the corresponding XUV (X) and NIR laser (L) beams. The
Rayleigh length $\zeta _{X,L}$ can be calculated as 
\begin{equation}
\zeta _{X,L}=\frac{w_{X,L}^{2}\omega _{X,L}}{2c},  \label{ZXL}
\end{equation}%
and is proportional to the frequency $\omega _{X,L}$, where $c$ is the speed of light.
The volume where ionization takes place is limited to the region
of overlap of the argon jet and the laser beam,
which we suppose to be symmetric at the waist, i.e., 
$-\zeta _{\text{jet}}\leq \zeta \leq \zeta _{\text{jet}}$.

The average over the focal volume of the angle-resolved photoelectron
spectrum is defined as

\begin{equation}
\left\langle \frac{dP}{\sin \theta dEd\theta }\right\rangle =\frac{1}{V}\int
\left( \frac{dP}{\sin \theta dEd\theta }\right) {dV,}
\label{def-avg-foc-vol}
\end{equation}%
where the $dP/(\sin \theta dEd\theta )$ is the electron emission
distribution calculated for the XUV and NIR intensities $I_{X}$ and $I_{L}$,
respectively. Considering the focal volume described above, Eq. (\ref%
{def-avg-foc-vol}) becomes%
\begin{equation}
\left\langle \frac{dP}{\sin \theta dEd\theta }\right\rangle =\frac{2\pi }{V}%
\int_{0}^{\infty }d\rho \ \rho \int_{-\zeta _{\text{jet}}}^{+\zeta _{\text{jet}}}d\zeta
\left( \frac{dP}{\sin \theta dEd\theta }\right) .  \label{avg-foc-vol}
\end{equation}
In order to calculate the average of Eq. (\ref{avg-foc-vol}), we
must integrate over the variables $\rho $ and $\zeta $ or, equivalently,
over the variables $I_{X}$ and $I_{L}$ through Eq. (\ref{IXL}). As the
calculations of $dP/(\sin \theta dEd\theta )$ for a single intensity are a
challenge by themselves, performing the average of Eq. (\ref{avg-foc-vol})
becomes computationally very expensive. For example, if we choose a grid of
$n$ points per each $I_{X}$ and $I_{L}$ intensity, we would need $n^{2}$
calculations of the distribution for the $n^{2}$ pairs of intensities $(I_{X},I_{L})$.
In the following we show how to overcome this difficulty with a valid physical criterion.

From TDSE and SFA calculations we have observed that for moderate laser
intensities the contribution to ionization due to the XUV and NIR laser pulses
are well separated in the energy domain: Whereas the single-photon XUV
ionization leads to ionization with final kinetic energy close
to $E\simeq \omega _{X}-I_{p}$, ionization due to the NIR pulse only
[setting  $\vec{F}_{X}=0$ in Eq. (\ref{hamiltonian}) and Eq. (\ref{Tif})] leads to
electrons with final kinetic energy less than twice the ponderomotive
energy $E\lesssim 2U_{p}$. If we focus on the emission due to the XUV pulse
around $E\simeq \omega _{X}-I_{p},$ the contribution from NIR
ionization is negligible provided $U_{p}\ll \omega _{X}-I_{p}$. With this in
mind and considering the first Born approximation for ionization by one XUV
photon, the ionization probability is proportional to the intensity of the
XUV field, i.e., $P \propto I_{X}$ in the ionization zone 
$-\zeta _{\text{jet}}\leq \zeta \leq \zeta _{\text{jet}}$. Moreover,
the NIR laser field affects the overall structure of the photoelectron
distribution with essentially no effect on the total ionization yield,
hence, we can say to a good degree of approximation that 
\begin{equation}
\left. \frac{dP}{\sin \theta dEd\theta }\right\vert _{I_{X},I_{L}}=\left( 
\frac{I_{X}}{I_{X0}}\right) \left. \frac{dP}{\sin \theta dEd\theta }%
\right\vert _{I_{X0},I_{L}}.  \label{property}
\end{equation}%
We have tested that Eq. (\ref{property}) is correct to a high level of accuracy
with TDSE and SFA calculations.
Since the evaluation for one single value of $I_{X0}$ is enough to calculate
the same for a different value $I_{X}$ through Eq. (\ref{property}), a grid in
the XUV intensity is not needed. Therefore, instead of $n^{2}$
values of the distribution [for the $n\times n$ values of $(I_{X},I_{L})$],
we need only $n$ different values of the distribution for the $n$ different
values of $I_{L}$.

\section{Experimental setup and results}
\label{experiment}

The experiments are carried out at the XUV time-delay-compensating monochromator
beamline described in detail elsewhere \cite{Eckstein15,Eckstein15a}.
In short, 1.5\,mJ pulses from a Ti:Sapp
laser system with durations of 25\,fs and repetition rate of 1\,kHz are used
to generate high-order harmonics in Ar. The XUV spectrum consists of a sequence
of odd harmonics of the driving laser field (800\,nm) spanning the photon energy
range from 10 to 50\,eV, approximately. The XUV beam is guided through the 
time-delay-compensating monochromator and one harmonic is selected. Thanks to the
compensation scheme the XUV pulse is not stretched by diffraction off the
monochromator gratings, but remains short with the duration close to that of the
driving pulse. The wavelength-selected XUV pulses are combined with the 800\,nm
NIR pulses using an annular mirror and are focused into the interaction region of
the velocity map imaging spectrometer (VMIS), which is capable of recording energy-
and angle-resolved spectra of photoemission from atoms and molecules. 

\begin{figure}
% Use the relevant command for your figure-insertion program
% to insert the figure file.
% For example, with the option graphics use
\resizebox{\columnwidth}{!}{%
  \includegraphics{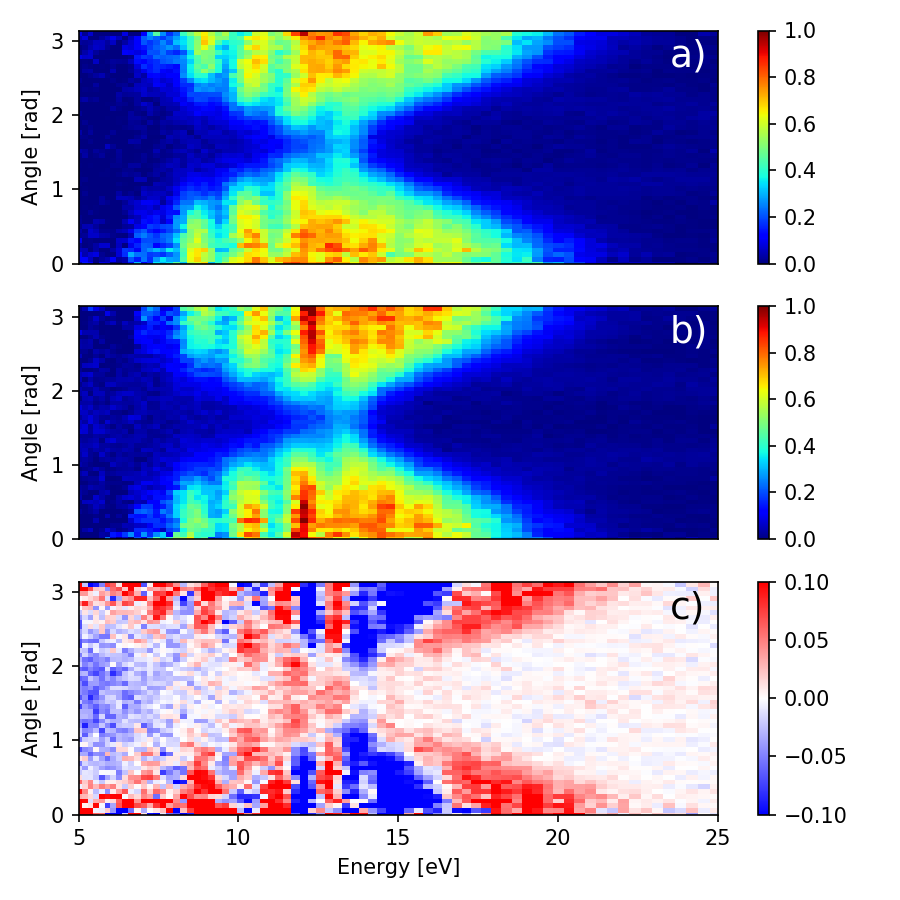}
}
% If not, use
%\vspace{5cm}       % Give the correct figure height in cm
\caption{Reconstructed energy-angle distributions of
laser-assisted photoemission spectra. The argon atoms are ionized by 29.6\,eV
XUV photons (harmonic 19) in the presence of the 800\,nm laser field with the
intensities of a) $14$\,TW/cm$^2$, b) $13$\,TW/cm$^2$. In (c) the difference map
for the two measurements of (a) and (b) is presented.
Red colors indicate the signal enhancement in the
velocity map image taken at higher laser field intensity, while blue colors
indicated the depletion.}
\label{fig:1}       % Give a unique label
\end{figure}

The LAPE images are recorded using the 29.6\,eV photons (harmonic 19) and
two different NIR field intensities determined by comparing the measured
spectra with the theoretical results of the next section.
VMIS records an Abel projection of the 3D photoelectron distributions.
The 3D distributions are recovered by Abel inversion of the raw experimental
images using the BASEX reconstruction algorithm.
Fig. \ref{fig:1} (a) and (b) shows the
reconstructed 2D slices through the full 3D photoelectron distributions
as false color energy-angle maps. Both maps show sets of vertical lines
corresponding to the main photoemission line (XUV-only)
and a set of sidebands with lower and higher kinetic energies.

% For one-column wide figures use

Although many sidebands can be observed in these experiments, even at the high
intensity of $14$ TW/cm$^2$ no modulations of the angular distributions (gross
structure) are visible in the images, although estimations predict that at these
laser fields the intracycle interference should play a role. 
From a preliminary study, we found first evidence that the absence
of the gross structure in these measurements is due to the averaging
over a range of NIR intensities in the focal volume \cite{Kubin13}.
In the present study, this finding is confirmed with calculated spectra,
averaged over different NIR intensities.
In the interaction zone of the VMIS both XUV and NIR beams have
comparable focus sizes. The waist radii are very difficult to determine, which
leads to uncertain XUV and IR laser intensities. Comparison to theoretical
calculations in the next section are compatible with intensities of about 14 TW/cm$^2$,
which corresponds to waist radii of about 100$\mu$m.
This is an efficient configuration for many pump-probe experiments,
but in the present case it results in ionization of Ar atoms taking
place at a range of NIR laser field intensities varying from the maximum
to almost zero across the focal volume.
Since positions of the angular features induced by intracycle interferences are
expected to depend on the intensity of the NIR field, the averaging
over the experimental focus washes out these structures. 

To recover the gross structure we subtract the distribution recorded in Fig.
\ref{fig:1} (b) from the corresponding of Fig. \ref{fig:1} (a), which are
taken at slightly different NIR field intensities. The result of this
subtraction is show in Fig. \ref{fig:1} (c) as a false-color map.
Subtracting one image from the other leaves predominantly contributions
due to higher peak laser intensity in Fig. \ref{fig:1} (a) and thus reveals
modulations in the emission angle corresponding to the intracycle interferences.
This method appears to be very practical for observing gross structures in
laser-assisted photoemission in the common case of matched focii 
of pump-probe XUV-NIR beams.

\section{Theoretical results and discussion}
\label{results}

\begin{figure}
\resizebox{\columnwidth}{!}{%
  \includegraphics{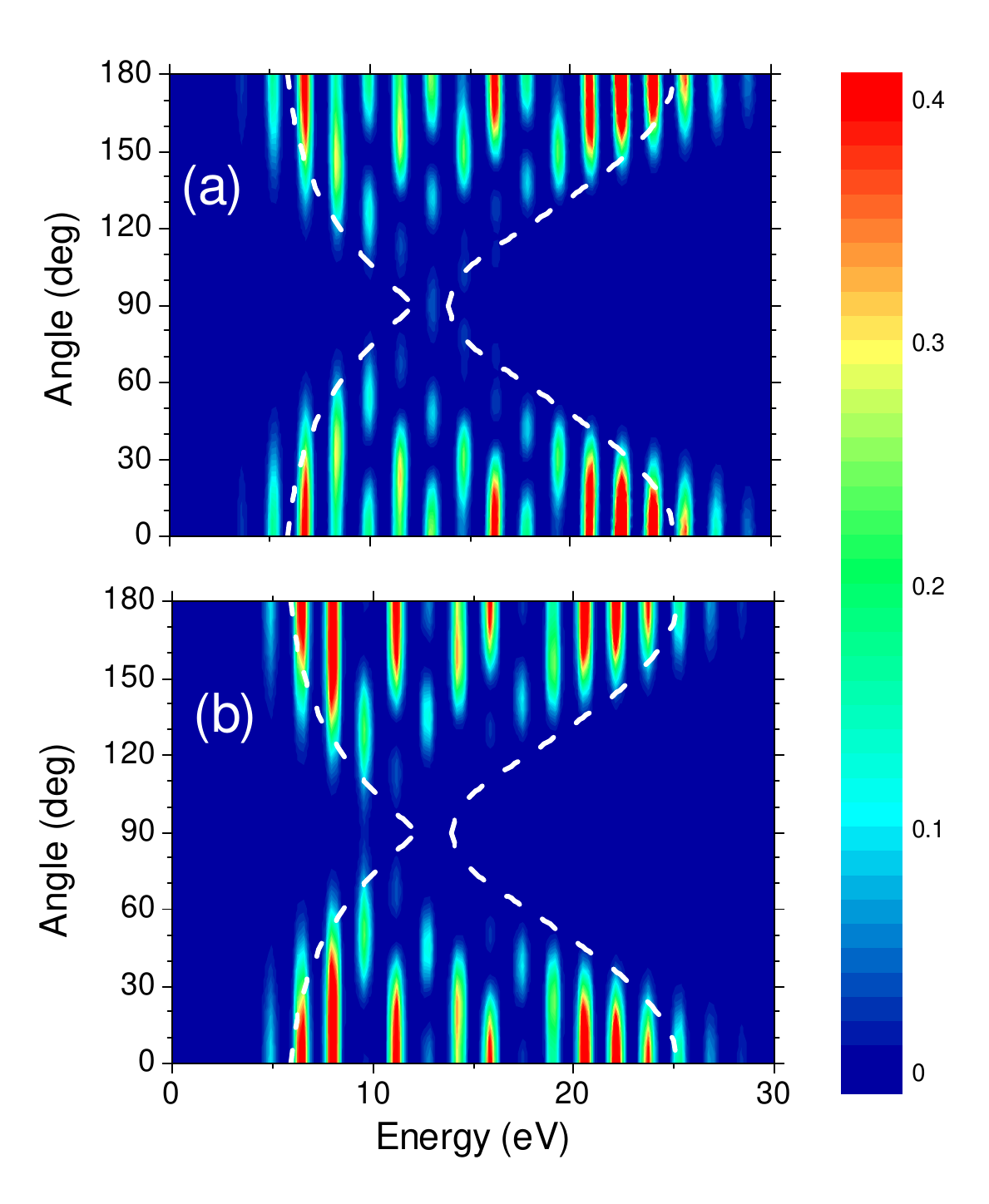}
}
\caption{Angle-energy probability distribution from Ar
subject to an XUV pulse of $\omega _{X}=1.0925$ ($29.7$ eV), duration $\tau
_{X}=5T_{L}=5\times (2\pi /\omega _{L})=546.4$ ($13.22$ fs) and a laser
pulse of frequency $\omega _{L}=0.0575$ ($1.56$ eV), duration $\tau
_{L}=7T_{L}=764.9$ ($18.5$ fs). The laser peak field is 
$F_{L0}=0.02$ ($I_{L}=1.4\times 10^{13}$ W/cm$^{2}$) and the XUV peak field
is $F_{X0}=0.01$ ($I_{X}=3.5\times 10^{12}$ W/cm$^{2}$).
(a) corresponds to TDSE calculations and (b) to SFA calculations.
Classical bounderies \cite{Gramajo18} are delimited by dashed lines.
}
\label{singI5c}       % Give a unique label
\end{figure}

We now turn to calculating LAPE spectra for the laser field parameters
used in the experiment.
To compare the different methods of calculation, TDSE and SFA,
described in Sec. \ref{sec:theory} to experimental measurements of 
Sec. \ref{experiment}, we consider a flattop envelope for the IR laser pulse
which is simple to model in the calculations 
(this assumption has no significant influence on the results). 
In this sense, the NIR laser field can be written as%
\begin{equation}
\vec{F}_{L}(t)=F_{L0}(t)\ \cos \left[ \omega _{L}\left( t-\frac{\tau _{L}}{2}%
\right) \right] \ \hat{z},  \label{IR-field}
\end{equation}%
where the envelope is given by%
\begin{equation}
F_{L0}(t)=F_{L0}\left\{ 
\begin{array}{ccc}
\frac{\omega _{L}t}{2\pi } & \mathrm{if} & 0\leq t\leq \frac{2\pi }{\omega
_{L}} \\ 
1 & \mathrm{if} & \frac{2\pi }{\omega _{L}}\leq t\leq \tau _{L}-\frac{2\pi }{%
\omega _{L}} \\ 
\frac{\left( \tau _{L}-t\right) \omega _{L}}{2\pi } & \mathrm{if} & \tau
_{L}-\frac{2\pi }{\omega _{L}}\leq t\leq \tau _{L}%
\end{array}%
\right.  \label{laser-envelope}
\end{equation}%
and zero otherwise. We define the XUV pulse as%
\begin{equation}
\vec{F}_{X}(t)=F_{X0}\ \sin ^{2}\left[ \frac{\pi }{\tau _{X}}\left( t-\frac{%
\tau _{L}-\tau _{X}}{2}\right) \right] \cos \left[ \omega _{X}\left( t-\frac{%
\tau _{L}}{2}\right) \right] \ \hat{z},  \label{XUV-field}
\end{equation}%
for $(\tau_L - \tau_X)/2 \leq t \leq (\tau_L + \tau_X)/2 $ and zero otherwise.
It is well known that XUV pulse shapes play a minor role in LAPE,
especially for long NIR pulses \cite{Duesterer13,Meyer10,Meyer12}. 
Therefore, we use a smooth sin$^{2}-$ envelope centered in the middle of
the NIR pulse. We consider that there is an integer number of optical
cycles in the XUV pulse, i.e., $N=\tau _{X}/2\pi \omega _{X}$.
The definitions of the NIR and XUV pulses in equations (\ref{IR-field})
and (\ref{laser-envelope}) assure a flattop vector potential $A(t)$
fulfilling the boundary conditions $A(0)=A(\tau _{L})=0$. In our calculations
pulse durations are defined as total widths, i.e., $\tau_{X,L}$, and not as FWHM.
We perform TDSE and SFA calculations for argon initially in a $3p$ state
($I_p \simeq 0.58 = 15.8$ eV) ionized by an XUV pulse of frequency
$\omega_{X}=1.0925$ ($29.73$ eV), duration $\tau _{X}=5T_{L}=5\times
(2\pi /\omega_{L})=546.4$ ($13.22$ fs) and an NIR laser pulse of frequency
$\omega _{L}=0.0575$ ($1.56$ eV), duration $\tau _{L}=7T_{L}=764.9$ ($18.5$ fs).
The angle- and energy-resolved electron distributions calculated via TDSE are
shown in Fig. \ref{singI5c} (a) for peak fields
$F_{L0}=0.02$ ($I_{L}=1.4\times 10^{13}$ W/cm$^{2}$).
Due to linear dependence of the electron yield on the XUV pulse intensity,
we have arbitrarily chosen $F_{X0}=0.01$ ($I=3.5\times 10^{12}$ W/cm$^{2}$).
The respective TDSE and SFA energy-angle distributions in Fig. \ref{singI5c}
(a) and (b) extend a bit beyond the classical boundaries drawn as dashed lines
\cite{Gramajo18} due to the quantum nature of the final wave function.
We observe that the electron emission is more probable along
the polarization axis, i.e. at $\theta =0{{}^\circ}$ and $180{{}^\circ}$.
According to Ref. \cite{Gramajo16}, the electron distribution for
emission parallel to the polarization axis is classically bounded, i.e,
\begin{equation}
\frac{(v_{0}-F_{L0}/\omega _{L})^{2}}{2}<E<\frac{(v_{0}+F_{L0}/\omega _{L})^{2}}{2},
\label{bound-par}
\end{equation} 
where $v_{0}^{2}/2=\omega _{X}-I_{p}$ corresponds to the electron kinetic
energy for ionization by an XUV pulse only. For argon ionization by the
electric fields with the laser parameters described above the energy domain along
the polarization axis must be classically bounded to $0.22$ ($6$ eV) $<E<0.92$
($25.1$ eV), where $v_{0}^{2}/2=0.51=13.9$ eV. According to Ref. 
\cite{Gramajo17}, in the perpendicular direction the electron kinetic energy
is classically restricted to 
\begin{equation}
\frac{\left[ v_{0}^{2}-\left( F_{L0}/\omega_{L}\right) ^{2}\right]}{2} <E<\frac{v_{0}^{2}/2}{2},
\label{bound-per}
\end{equation}
which corresponds to $0.45$ ($12.3$ eV) $<E<$ $0.51$ ($13.9$ eV).

The vertical iso-energy stripes in Fig. \ref{singI5c} (a) corresponds to
the sidebands arising from the absorption of one XUV photon followed 
by the absorption/emission of $n$ NIR photons
in agreement with the conservation of energy equation%
\begin{equation}
E_{n}=n\omega _{L}+\omega _{X}-I_{p}-U_{p}.  \label{En}
\end{equation}%
It is worth noticing that the energy of the XUV mainline together with
all SBs are slightly shifted down by the ponderomotive energy of the NIR laser
$U_{p}=(F_{L0}/2\omega_{L})^{2}=0.03$ ($0.8$ eV).
For perpendicular emission ($\theta =90{{}^\circ}$),
the even sidebands are missing due to the selection rule for angular momentum
in agreement with theoretical predictions \cite{Gramajo17}.
We can also observe in Fig. \ref{singI5c} (a) that the
sidebands are modulated by angle- and energy-dependent stripes whose origin stem 
from the interference of electron emission within the same optical cycle,
i.e., the intracycle interference pattern (see \cite{Gramajo16,Gramajo17,Gramajo18}).
In Fig. \ref{singI5c} (b) we show the energy-angle distribution calculated
within the SFA. The resemblance to the TDSE distribution of Fig. \ref{singI5c} (a)
is outstanding, which tells us about the minor role of the Coulomb potential
in the photoionization process \cite{Arbo10b}.

\begin{figure}
\resizebox{\columnwidth}{!}{%
  \includegraphics{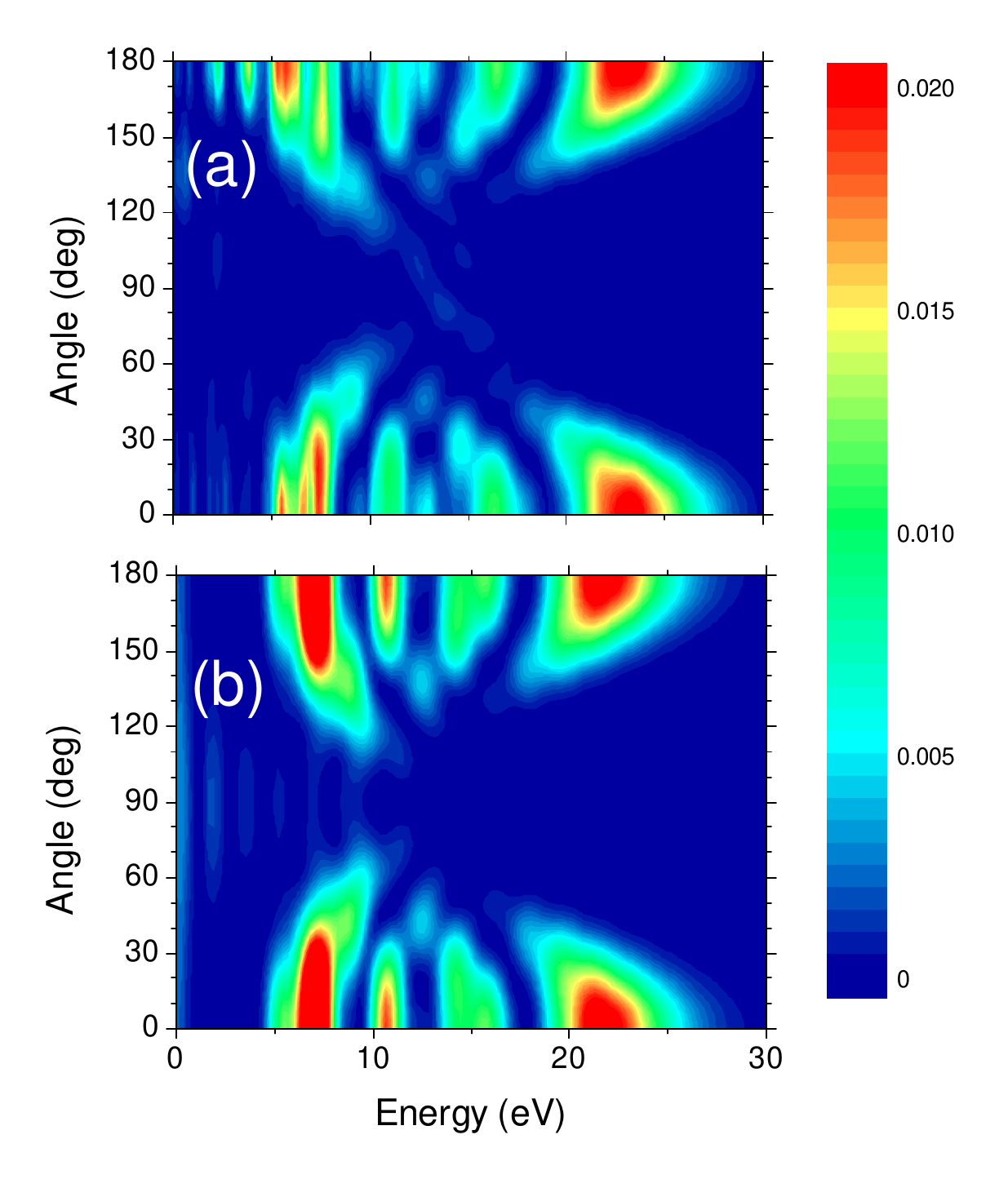}
}
\caption{Angle-energy probability distribution from Ar
subject to flat-top XUV pulse of $\omega _{X}=1.0925$ ($29.73$ eV), duration $\tau
_{X}=T_{L}= (2\pi /\omega _{L})=109.3$ ($2.64$ fs) and a laser
pulse of frequency $\omega _{L}=0.0575$ ($1.56$ eV) and duration $\tau
_{L}=7T_{L}=764.9$ ($18.5$ fs). The laser peak field is 
$F_{L0}=0.02$ ($I_{L}=1.4\times 10^{13}$ W/cm$^{2}$) and the XUV peak field
is $F_{X0}=0.01$ ($I_{X}=3.5\times 10^{12}$ W/cm$^{2}$).
(a) corresponds to TDSE calculations and (b) to SFA calculations.
}
\label{singI1c}       % Give a unique label
\end{figure}

The SBs in Fig. \ref{singI5c} are modulated by a gross
structure which depends on both energy and angle. 
In order to prove that this structure stems from the intracycle
interference of electron trajectories released
within the same optical laser cycle, we have
performed a calculations for an XUV pulse duration of one optical cycle,
i.e., $\tau_{X}=T_{L}= (2\pi /\omega _{L})=109.3$ ($2.64$ fs) considering a
flat-top envelope instead of the $\sin^2$ envelope in Eq. (\ref{XUV-field}).
In Fig. \ref{singI1c} we observe that as ionization takes place essentially within
the only one cycle of the NIR field, no SBs are formed, and only the intracycle
interference structures remain visible in the maps. When we compare 
Fig. \ref{singI1c} with Fig. \ref{singI5c}, we observe that the modulating stripes
for the five-cycle XUV pulse in Fig. \ref{singI5c} look essentially the same as 
the intracycle structure for the $\tau _{X}=T_{L}$ calculation in Fig. \ref{singI1c}.
Again, SFA calculations in Fig. \ref{singI1c} (b) are in good agreement
with TDSE calculations in Fig. \ref{singI1c} (a).

\begin{figure}
\resizebox{\columnwidth}{!}{%
  \includegraphics{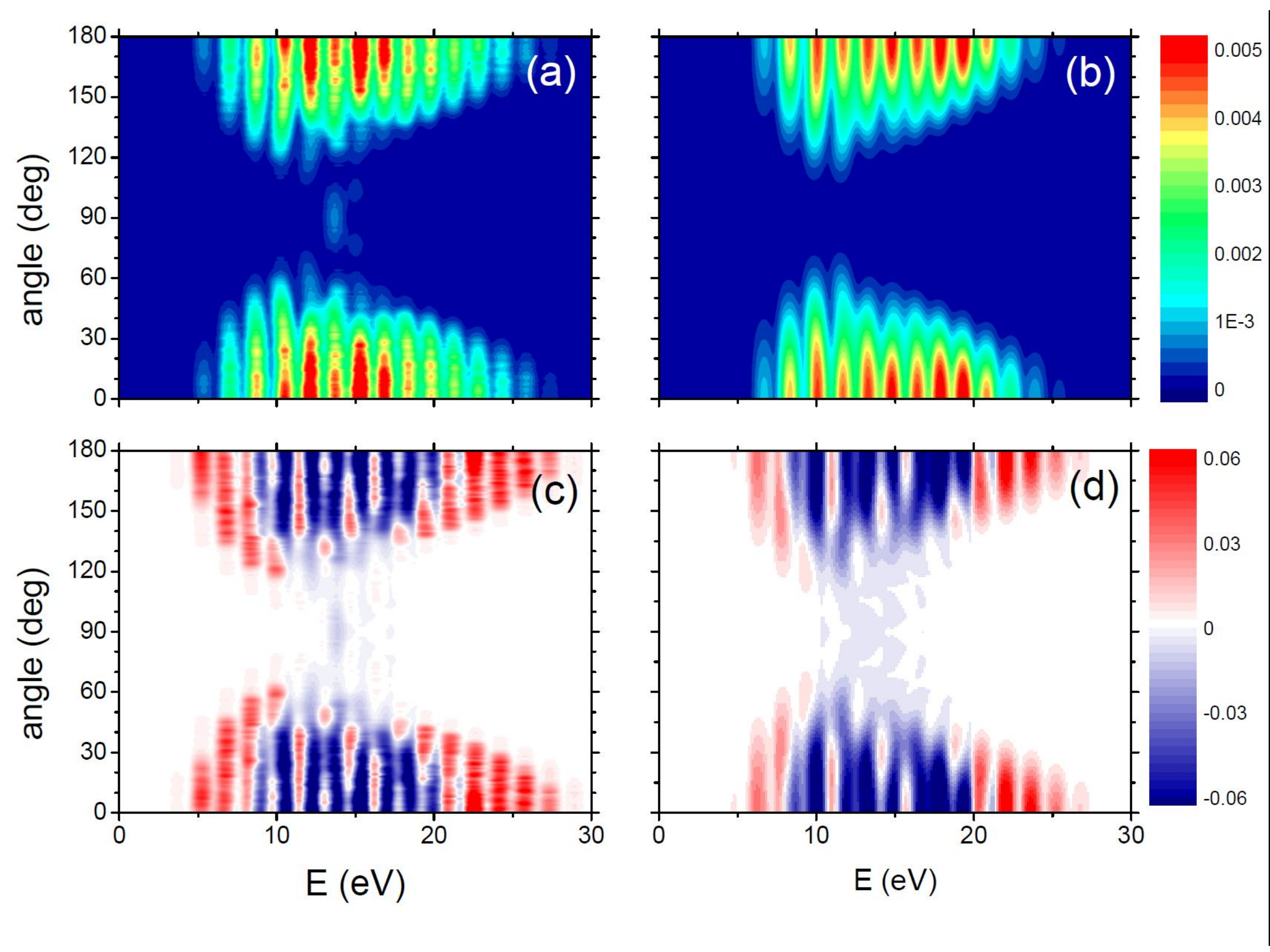}
}
\caption{Angle-energy probability distribution from Ar(3p)
subject to an XUV pulse of peak field is $F_{X0}=0.01$ ($I_{X}=3.5\times 10^{12}$ W/cm$^{2}$),
$\omega _{X}=1.0925$ ($29.73$ eV), duration 
$\tau_{X}=5T_{L}=5\times (2\pi /\omega _{L})=546.4$ ($13.22$ fs)
and a laser pulse of frequency $\omega _{L}=0.0575$ ($1.56$ eV),
duration $\tau_{L}=7T_{L}=764.9$ ($18.5$ fs), and laser peak field
$F_{L0}=0.04$ ($I_{L}=5.6\times 10^{13}$ W/cm$^{2}$).
(a) and (b) distributions are averaged over the focal volume [Eq. (\ref{avg-foc-vol})],
(c) and (d) corresponds to the difference between the distributions calculated in (a) and (b)
and the ones in with $F_{X0}=0.0095$ ($I_{X}=3.2\times 10^{12}$ W/cm$^{2}$) and
$F_{L0}=0.019$ ($I_{L}=1.27\times 10^{13}$ W/cm$^{2}$).
(a) and (c) corresponds to TDSE calculations and (b) and (d) to SFA calculations.
Calculations were smoothed using an energy window of about $1$ eV and an angle
of about $5^\circ$.}
\label{avgp02}       % Give a unique label
\end{figure}

As discussed in Sec. \ref{experiment}, if we want to compare our simulations
with the experiment, we need to average the electron spectra over the
experimental intensity distribution in the focal volume 
(see Sec. \ref{sec:theory}.3). We assume a Gaussian profile of the NIR laser beam
with a waist radius of $w_{L}=0.1$ mm corresponding to a Rayleigh length of
$\zeta _{L}=10$ mm and also a Gaussian profile for the XUV beam with the same
radius $w_{X}=w_{L}$ and corresponding Rayleigh length $\zeta _{X}=190$ mm
[see Eq. (\ref{ZXL})]. We suppose an atomic beam with aperture $2\zeta
_{\text{jet}}=10 $ mm. To reduce computation time, averaging is done
by performing calculations for twenty values of pairs $(I_{X,i},I_{L,i})$
with a constant ratio $I_{X,i}/I_{L,i}$ and then extrapolating
for different values of $I_{X}$, according to Eq. (\ref{property}).
In Figs. \ref{avgp02} (a) and (b) we see the result of the averaging over
the focal volume of the respective TDSE and SFA calculated distributions
according to Eq. (\ref{avg-foc-vol}) for an XUV peak field
in the focus of $F_{X0}=0.01$ corresponding to an intensity
$I_{X}=3.5\times 10^{12}$ W/cm$^{2}$ and a NIR laser peak field in the focus of 
$F_{L0}=0.02$ corresponding to an intensity $I_{L}=1.4\times 10^{13}$ W/cm$^{2}$,
the laser and XUV frequencies and durations are the same as in Fig.
\ref{singI5c}. We observe that the general features of the distribution are
similar to the non-averaged distributions in Fig. \ref{singI5c} (a) except for
the smoothing and disappearing of the intracycle interference pattern.
This lack of intracycle interference pattern agrees with the experimental
observations [see Fig. \ref{fig:1} (a) and (b)].
In Fig. \ref{avgp02} (a) and (b) we observe that the probability distribution in the 
direction parallel to the polarization axis extends from about $5$ to $25$ eV
approximately, very similar to the experimental data in Fig. \ref{fig:1} (a) and (b).
Besides, for emission perpendicular to the polarization axis, the maximum lies about
$14$ eV in both calculations [in Fig. \ref{avgp02} (a) and (b)] and measurements 
[in Fig. \ref{fig:1} (a) and (b)].

In order to recover the intracycle interference pattern,
we repeat the procedure performed for the experimental data
and calculate a difference map subtracting two similar intensity-averaged
distributions with peak intensities of the NIR laser
$I_L=1.41\times 10^{13}$ W/cm$^{2}$ and $I_L=1.27\times 10^{13}$ W/cm$^{2}$
corresponding to electric fields of $F_{L0}=0.02$ and $F_{L0}=0.019$,
respectively. As the total measured ionization yield is unknown,
we normalize the two calculated averaged electron-angle distributions
before subtracting. Therefore, the difference shows positive and negative values
in Fig. \ref{avgp02} (c) and (d) for TDSE and SFA, respectively.
Even though the distribution in Fig. \ref{avgp02} (c) and (d) cannot be thought of as
a probability distribution due to the existence of negative values,
the intracycle structure pattern can be easily identified in the figure
(compare to single intensity non-intensity-averaged probability distribution in
Fig. \ref{singI5c}). There is a strong similarity between the intracycle interference
pattern of Figs. \ref{singI1c} (a) and (b) and the difference map of Figs. \ref{avgp02}
(c) and (d) for the respective TDSE and SFA calculations. To account for experimental
resolution we have smoothed the calculations using an energy window of about $1$ eV
and an angle of about $5^\circ$. The intracycle interference pattern is recovered,
as can be easily seen by comparing it to Figs. \ref{singI5c} (a) and (b).
Besides, from comparing Fig. \ref{singI1c} (b) to (a)
we can say that the SFA reproduce the TDSE quite accurately. 
n both the TDSE and the SFA calculations the interference structure can be retrieved
with the subtraction procedure used for the treatment of the experimental data.
Therefore we can state that our calculations both reproduce the experiment remarkably well.

\section{Conclusions}
\label{conclusions}
We performed experiments on laser-assisted XUV ionization of argon atoms and
recorded energy- and angle-resolved photoelectron spectra.
We see that the intracycle interference pattern is smeared out and cannot
be directly observed after accounted for the average over the focal volume.
However, by subtracting two angle-resolved photoelectron spectra
for slightly different NIR laser intensities the intracycle interference pattern is
recovered. We have performed TDSE \textit{ab initio} and SFA
calculations and averaged them over the focal volume of the XUV and NIR laser
pulses to compare with experiments. Both TDSE and SFA are in good agreement
corresponding well to the experiment.  By performing simulations using a
flattop XUV pulse that comprises only one optical laser cycle,
we observed a clear intracycle interference pattern very similar to the theoretically
and experimentally extracted by subtracting distributions for slightly different
laser intensities. We have also shown that theoretical calculations of LAPE are a
useful tool to determine the elusive experimental NIR laser intensity.

We have retrieved intracycle interference fringes in electron emission
produced by atomic argon ionization subject to an XUV pulse in the presence
of a strong near-infrared laser pulse with both pulses linearly polarized
in the same direction. 
%
% The section below may be edited at your convenience to acknowledge 
% each author's contribution to the manuscript.
% You may remove it if you are a single author.
%
\section{Acknowledgements}

We are grateful to N. Kabachnik for fruitful discussions and early support.
Work supported by bilateral Argentine-German grant CONICET-DAAD of 2015,
the grant of Deutsche Forschungsgemeinschaft (KO 4920/1-1),
and by CONICET PIP0386, PICT-2016-0296 and PICT-2014-2363 of ANPCyT (Argentina).

\section{Authors contributions}
J. Hummert, M. Kubin, and O. Kornilov performed the experiment,
and S. D. L\'{o}pez and D. G. Arb\'{o} performed the theoretical calculations.
All the authors were involved in the preparation of the manuscript and have read and approved
the final manuscript.
%
% BibTeX users please use
%\bibliographystyle{plain}
\bibliographystyle{unsrt}
\bibliography{biblio-diego}
%
% Non-BibTeX users please use
%\begin{thebibliography}{}
%
% and use \bibitem to create references.
%
%\bibitem{RefJ}
% Format for Journal Reference
%Author, Journal \textbf{Volume}, (year) page numbers.
% Format for books
%\bibitem{RefB}
%Author, \textit{Book title} (Publisher, place year) page numbers
% etc
%\end{thebibliography}

\end{document}